\documentclass[12pt,a4paper,twocolumn]{article}
\usepackage{amsfonts}
\usepackage{graphicx}
\usepackage{amsmath}
\usepackage{a4}

\input{tcilatex}
\begin{document}

\title{ Equivalent random propagation time for coaxial cables}
\author{B. Lacaze \\
Tesa, 14/16 Port St-Etienne 31000 Toulouse France}
\maketitle

\begin{abstract}
Propagation of monochromatic electromagnetic waves in free space results in
a widening of the spectral line. On the contrary, propagation preserves
monochromaticity in the case of acoustic waves. In this case, the
propagation can be modelled by a linear invariant filter leading to
attenuations and phase changes. Due to the Beer-Lambert law, the associated
transfer function is an exponential of power functions with
frequency-dependent parameters.

In recent papers, we have proved that the acoustic propagation time can be
modelled as a random variable following a stable probability distribution.
In this paper, we show that the same model can be applied to the propagation
in coaxial cables.

\textit{keywords: }random propagation time, stationary process, stable
probability distributions, coaxial cables.
\end{abstract}

\section{Introduction}

1) Coaxial cables are characterized by four parameters $R,C,L,G$ and a
length $l\ $with respective units metre ($m$), ohm ($\Omega $), farad ($F$),
henry by metre ($H/m$). The Beer-Lambert law and elementary properties of
circuits lead to the following tranfer function%
\begin{equation}
H_{l}\left( \omega \right) =e^{-l\sqrt{\left( R+i\omega L\right) \left(
G+i\omega C\right) }}.
\end{equation}%
In practice, the dielectric loss is generally negligible compared to $\omega
C$ and $R$ depends on the frequency $\omega /2\pi $ \cite{Jin}.
Consequently, the transfer function simplifies as follows: 
\begin{equation*}
H_{l}\left( \omega \right) =e^{-l\left( im\omega +a\sqrt{\omega }\left(
1+i\right) \right) },\text{ for }\omega >0
\end{equation*}%
where the parameters $m$ and $a$ characterize the coaxial cable in the
considered frequency band. This formula is generalized to $\omega \in 
\mathbb{R}$:%
\begin{equation}
H_{l}\left( \omega \right) =e^{-l\left( im\omega +a\sqrt{\left\vert \omega
\right\vert }\left( 1+i\text{sgn}\omega \right) \right) }
\end{equation}%
using the relation $H_{l}\left( \omega \right) =H_{l}^{\ast }\left( -\omega
\right) $ since a real input must yield a real output. The exponential
expression $H_{l}\left( \omega \right) =e^{-l\gamma \left( \omega \right) }$
is in accordance with the Beer-Lambert law and thus to the relation relating
filters in series:%
\begin{equation*}
H_{l+l^{\prime }}\left( \omega \right) =H_{l}\left( \omega \right)
H_{l^{\prime }}\left( \omega \right) .
\end{equation*}%
The relation $\left( 2\right) $ means that the monochromatic wave $%
e^{i\omega t}$ at the cable input is transformed in the monochromatic wave $%
H_{l}\left( \omega \right) e^{i\omega t}$ at the distance $l,$ assuming that
the cable is semi-infinite or well-matched.

2) This result can be compared with propagation in other media. Propagation
of electromagnetic waves in free (i.e. not guided) medium often leads to a
widening of the spectral line. Backscattering of monochromatic waves on
trees in the X-band (8 GHz) results in a power spectrum with a bandwidth
around few tens Hz \cite{Laca8}, \cite{Nara}. The same behavior has been
observed for laser propagation in the atmosphere \cite{Laca3}, \cite{Parr}.
In sea water, radar backscattering leads to a mixing of Gaussian spectra
depending on the polarization and with Doppler shift \cite{Laca9}, \ \cite%
{Walk}. The same phenomena are found in wind profilers \cite{Laca3}, \cite%
{Bals}. We have known also for a long time that emissions or absorptions of
sky light lead to line broadening and Doppler shift mainly explained by star
composers and movements \cite{Laca7}. In all these practical situations, the
observed spectral widening changes a line spectrum into a continuous one. To
my knowledge and for small powers, spectral widening has not been noticed in
coaxial cables. The cable behaves like a linear invariant filter subject to
the Beer-Lambert law which results to an exponential expression of its
transfer function. Consequently, for a coaxial cable, a pure monochromatic
wave remains a pure monochromatic wave. The amplitude and the phase are two
functions of the frequency $\omega /2\pi $ defined by formula $\left(
2\right) $.

3) The same behavior is observed for acoustic propagation. The received wave
is monochromatic like the transmitted wave. In the atmosphere or in the sea
water, the wave attenuation expresses as exp$\left[ -al\omega ^{2}\right] $,
where the parameter $a$ is medium-dependent and $l$ is the distance \cite%
{Laca6}. The celerity $c$ of the wave is constant with respect to the
frequency. Actually, tables giving the influence of the temperature, the
salinity and other physical parameters are very detailed. To our knowledge
there exists no table giving the dependence between $c$ and~$\omega $ since
they are usually assumed independent. However, ultrasonic waves have a
different behavior. The crossing of media like biological tissues, liquids
like castor oil (mimicking tissues), egg yolk or many other substances on
small distances (of the order of the centimetre for instance) are weakened
like exp$\left[ -al\omega ^{b}\right] $ with $0<b<2$ \cite{Laca4}, \cite%
{Laca5}, \cite{Szab}. Moreover, for $b\neq 2,$ the celerity is function of $%
\omega $. For $b\neq 1,$ the complex gain $H_{l}\left( \omega \right) $ of
the equivalent filter verifies$:$%
\begin{equation}
\begin{array}{c}
H_{l}\left( \omega \right) = \\ 
\exp \left[ -ilm\omega -al\left\vert \omega \right\vert ^{b}\left( 1+i\frac{%
\omega }{\left\vert \omega \right\vert }\tan \frac{\pi b}{2}\right) \right]%
\end{array}%
\end{equation}%
and for $b=1:$%
\begin{equation*}
\begin{array}{c}
H_{l}\left( \omega \right) = \\ 
\exp \left[ -ilm\omega -al\left\vert \omega \right\vert \left( 1-i\frac{%
\omega }{\left\vert \omega \right\vert }\frac{2\ln \left\vert \omega
\right\vert }{\pi }\right) \right]%
\end{array}%
\end{equation*}%
For instance, the value $b=1$ often characterizes biological tissues or
evaporated milk. $b=1.66$ is used for castor oil up to 300MHz with a very
good accuracy, $b$ is around 1.5 for egg yolk and $b=0.5$ is for brass tubes
in low frequency (400-2400Hz) \cite{Maso}... Equations $\left( 3\right) $
are in accordance with the "near local Kramers-Kronig theory" of Szabo \cite%
{Szab}. We note that, in the particular case where $b=2,$ the celerity of
the wave is the constant $m$ ($\tan \frac{\pi b}{2}=0)$ according to Eq.$%
\left( 3\right) .$

Propagation in coaxial cables verifies $\left( 2\right) ,$ i.e. $\left(
3\right) $ with $b=1/2.$ This result appears as a paradox: this behavior is
similar to the case of acoustic waves (preserved monochromaticity) and not
to electromagnetic wave propagation in free space (widened power spectrum).

4) After these considerations about physical modelling, some probabilistic
models can be discussed. The probability distribution of the random variable
(r.v.) $X$ belongs to the set of stable distributions when its
characteristic function (c.f.) is defined by:%
\begin{equation}
\begin{array}{c}
\text{E}\left[ e^{-i\omega X}\right] = \\ 
\exp \left[ -i\alpha \omega -al\left\vert \omega \right\vert ^{b}\left( 1-ic%
\frac{\omega }{\left\vert \omega \right\vert }\tan \frac{\pi b}{2}\right) %
\right]%
\end{array}%
\end{equation}%
when $b\neq 1$ and, when $b=1$%
\begin{equation*}
\begin{array}{c}
\text{E}\left[ e^{-i\omega X}\right] = \\ 
\exp \left[ -i\alpha \omega -al\left\vert \omega \right\vert \left( 1+ic%
\frac{\omega }{\left\vert \omega \right\vert }\frac{2\ln \left\vert \omega
\right\vert }{\pi }\right) \right] .%
\end{array}%
\end{equation*}%
The set of real parameters $\left( \alpha ,a,b,c\right) $ must verify the
conditions $a>0,0<b\leq 2,\left\vert c\right\vert \leq 1$ \cite{Luka}, \cite%
{Samo}$.$ Stable distributions generalize the Central-Limit theorem to r.v.
with infinite variance. The complex gains $H_{l}\left( \omega \right) $
defined by $\left( 3\right) $ can be identified with the subset of functions
in $\left( 4\right) $ under the condition $c=-1.$ Moreover a random process $%
\mathbf{A=}\left\{ A\left( t\right) ,t\in \mathbb{R}\right\} ,$ such that
the r.v $A\left( t\right) $ and $A\left( t\right) -A\left( t-\tau \right) $
follow stable distributions with $b=\frac{1}{2},c=-1$ and $b=\frac{1}{2},c=0$
respectively, can be derived.

As shown in the next section, the attenuations and phase changes in coaxial
cables can be explained by random propagation times. Numerical values come
from the data sheet of Belden 8281. This class of cables has been studied in
several theses on equalization (for instance see \cite{Chen}, \cite{Hadg}).

\section{Random propagation times}

1) Let $\mathbf{A}_{l}\mathbf{=}\left\{ A_{l}\left( t\right) ,t\in \mathbb{R}%
\right\} $ be a random process with the following one-dimensional
characteristic function 
\begin{equation}
\text{E}\left[ e^{-i\omega A_{l}\left( t\right) }\right] =e^{-l\left(
im\omega +a\sqrt{\left\vert \omega \right\vert }\left( 1+i\text{sgn}\omega
\right) \right) }.
\end{equation}%
This formula corresponds to a stable probability distribution (4) with
parameters $\left( lm,la,\frac{1}{2},-1\right) $ along with the transfer
function $\left( 2\right) $ which defines a coaxial cable. Now, let $\mathbf{%
Z}_{l}\mathbf{=}\left\{ Z_{l}\left( t\right) ,t\in \mathbb{R}\right\} $
denote the random process defined by:%
\begin{equation}
Z_{l}\left( t\right) =e^{i\omega _{0}\left( t-A_{l}\left( t\right) \right) }.
\end{equation}%
$Z_{l}\left( t\right) $ is the output of a device with input $e^{i\omega
_{0}t}$ $(\omega _{0}>0)$ and subjected to a random propagation time $%
A_{l}\left( t\right) .$ Let assume that $\mathbf{A}_{l}$ is stationary in
the sense where%
\begin{equation*}
\phi _{l}\left( \omega ,\tau \right) =\text{E}\left[ e^{-i\omega \left(
A_{l}\left( t\right) -A_{l}\left( t-\tau \right) \right) }\right]
\end{equation*}%
is independent of $t$ and sufficiently regular. The process $\mathbf{Z}_{l}$
can be split in two additive terms \cite{Laca1}: 
\begin{equation}
\mathbf{Z}_{l}=\mathbf{G}_{l}+\mathbf{V}_{l}
\end{equation}%
where $\mathbf{G}_{l}=\left\{ G_{l}\left( t\right) ,t\in \mathbb{R}\right\} $
is defined by 
\begin{equation}
G_{l}\left( t\right) =e^{i\omega _{0}\left( t-lm\right) -al\sqrt{\omega _{0}}%
\left( 1+i\right) }.
\end{equation}%
The process $\mathbf{V}_{l}$ defined by $\left( 7\right) $ and $\left(
8\right) $ is zero-mean and stationary with autocorrelation function%
\begin{equation}
\begin{array}{c}
\text{E}\left[ V_{l}\left( t\right) V_{l}^{\ast }\left( t-\tau \right) %
\right] = \\ 
e^{i\omega _{0}\tau }\left[ \phi _{l}\left( \omega _{0},\tau \right) -e^{-2al%
\sqrt{\omega _{0}}}\right] .%
\end{array}%
\end{equation}

2) If we identify the wave with the model $\left( 6\right) ,$then $\mathbf{G}%
_{l}$ is the wave measured at distance $l.$ Because the power of \textbf{Z}$%
_{l}$\textbf{\ }is constant, $\mathbf{V}_{l}$ represents the losses in the
cable and in the medium up to the distance $l$. This quantity is not
measured by practical devices and is probably outside the observed frequency
band. From $\left( 9\right) $ the $\mathbf{V}_{l}-$power spectrum depends on
the probability distribution of $A_{l}\left( t\right) -A_{l}\left( t-\tau
\right) .$ In the appendix, we prove that it is possible to construct
processes \textbf{A}$_{l}$ which fulfill the following conditions:

a) the r.v. $A_{l}\left( t\right) $ possesses the stable distribution
defined by $\left( 5\right) $ with arbitrary parameters $m,a$

b) the r.v. $A_{l}\left( t\right) -A_{l}\left( t-\tau \right) $ possesses
the stable distribution defined by $\left( 4\right) $ with $\alpha =c=0,b=%
\frac{1}{2}$

c) the construction can be made so that the $\mathbf{V}_{l}-$power in any
frequency band $\left( \omega _{0}-b,\omega _{0}+b\right) $ is arbitrarily
small.

The last property explains why the model of propagation \textbf{Z}$_{l}$ \
defined by $\left( 6\right) $ fulfills the theorem of the energy balance
though the measured wave is an attenuated (and delayed) replica of the
transmitted wave. The process \textbf{V}$_{l}$ is the quantity lost and
dissipated by the medium at frequencies far from the transmitted wave
frequency.

3) In the coaxial cable framework $\mathbf{G}_{l}$ is the received wave at
the distance $l.$ For the Belden 8281 cable in the band (1,1000MHz) we have 
\cite{Chen}%
\begin{equation*}
m=\sqrt{LC}=52.10^{-10}\text{s.m}^{-1}
\end{equation*}%
which leads to a wave celerity equal to $2.10^{\text{8}}$m.s$^{-1}$ i.e.
66\%\ of the light celerity in vacuum. Moreover, from the same source and in
the usual system 
\begin{equation*}
a=39.10^{-8}.
\end{equation*}%
From $\left( 8\right) ,$ the term $al/\sqrt{\omega _{0}}$ is an extra delay
for the wave. We have to compare the nominal delay $ml$ with the variable
delay $al/\sqrt{\omega _{0}}.$ At 100MHz, the extra-delay is smaller than
1\% of the nominal delay.

\section{Remarks}

1) The theory of stable probability distributions reveals the following
interesting property \cite{Luka}, \cite{Samo}. Among the c.f. $\left(
4\right) ,$ the only one-sided probability densities are defined by the
parameters values $c=\pm 1,b<1.$ $c=1$ is for a one-sided to the left and $%
c=-1$ for an one-sided to the right (the probability density is 0 at the
left of the origin point when $\alpha =0).$ This is equivalent to the
causality of the filter defined by the transfer function $\left( 2\right) $.
For $b\geq 1$ or $b<1,c\neq \pm 1,$ this property does not hold. However,
among the $c-$parameter values, the value $-1$ is the best one, because it
minimizes the probability at the left of\ the origin point. It is indeed the
value $c=-1$ which has been taken in $\left( 4\right) $ to verify $\left(
3\right) .$ Moreover, the imaginary part in the exponential of $\left(
2\right) $ is dominated by the term $-ilm\omega _{0}$ in real cases$.$
Though the term $-ila\sqrt{\left\vert \omega _{0}\right\vert }$ seems
negligible with respect to $-iml\omega _{0}$, its influence on the
corresponding impulse response is strong.

Finally, the conditions of the Kramers-Kronig relations are verified by $%
K\left( z\right) $ defined by%
\begin{equation*}
\begin{array}{c}
K\left( z\right) =\exp \left[ -alz^{b}\left( 1-i\tan \frac{\pi b}{2}\right) %
\right] \\ 
al>0,0<b<1%
\end{array}%
\end{equation*}%
when $z^{b}=\rho ^{b}e^{ib\theta }$ with $\rho >0,0<\theta <\pi $ in the
upper plane. On the real axis, $K\left( \omega \right) =H_{l}\left( -\omega
\right) $ when $m=0.$

2) For frequencies larger than 1000MHz, the model $\left( 6\right) $ is no
longer sufficient for the Belden 8281 cable (and from some other frequencies
for other cables). It has to be changed in 
\begin{equation}
H_{l}\left( \omega \right) =e^{-l\left( im\omega +a\sqrt{\left\vert \omega
\right\vert }\left( 1+i\text{sgn}\omega \right) +\gamma \left\vert \omega
\right\vert \right) }
\end{equation}%
with $\gamma >0.$ The new term in $\gamma \left\vert \omega \right\vert $
takes into account dielectric losses \cite{Jin}. We know that $e^{-\lambda
\left\vert \omega \right\vert },\lambda >0,$ is the c.f. of the Cauchy
distribution. Consequently $\left( 10\right) $ corresponds to the
convolution of the stable distribution $\left( lm,la,\frac{1}{2},-1\right) $
with a Cauchy distribution with parameter $\gamma l.$ Though the Cauchy
distribution is stable (with parameters $(0$,$l\gamma ,1,0))$ the result no
longer defines a stable distribution.

3) The transfer function $H\left( \omega \right) $ is the Fourier transform
of the impulse response $h\left( t\right) :$%
\begin{equation*}
H\left( \omega \right) =\int_{-\infty }^{\infty }h\left( t\right)
e^{-i\omega t}dt.
\end{equation*}%
This definition is coherent with the input-output relation%
\begin{equation*}
y_{out}\left( t\right) =\int_{-\infty }^{\infty }h\left( u\right)
y_{in}\left( t-u\right) du
\end{equation*}%
Equivalently $H\left( \omega \right) e^{i\omega t}$ is the output when $%
e^{i\omega t}$ is the input and/or $h\left( t\right) $ is the output when $%
\delta \left( t\right) $ (the "Dirac function") is the input.

In probability calculus a c.f. $\psi \left( \omega \right) $ is the Fourier
transform of a probability density $f\left( t\right) $ (if it exists) in the
sense%
\begin{equation*}
\psi \left( \omega \right) =\int_{-\infty }^{\infty }f\left( t\right)
e^{i\omega t}dt.
\end{equation*}%
To identify a transfer function with a characteristic function it is
necessary to change $e^{i\omega t}$ into $e^{-i\omega t}$ in the last
equation (compare formulas $\left( 4\right) $ in this paper with formula
5.7.19 in Lukacs \cite{Luka}).

4) The probability density of $A_{1}\left( t\right) $ (and of $A_{l}\left(
t\right) $ whatever $l)$ is given from $\left( 5\right) ,\left( 11\right)
,\left( 12\right) .$ For the Belden 8281 coaxial cable with $a=39.10^{-8},$
we have approximately%
\begin{equation*}
\text{Pr}\left[ A_{1}\left( t\right) -m>8.10^{-10}\right] \simeq 0.01
\end{equation*}%
to be compared with $m=52.10^{-10}$s.m$^{-1}.$ The mode is close to 5.10$%
^{-14}$ and it is well-known that this distribution is heavy-tailed.

\section{Conclusion}

In circuit theory, a coaxial cable is defined by a set $\left(
R,C,L,G\right) $ where $\left( R,L\right) $ and $\left( C,G\right) $
represent series and parallel components. The equivalent circuit highlights
the linear functions $\left( R+i\omega L\right) $ and $\left( G+i\omega
C\right) $ of the frequency $\omega /2\pi .$ Actually, it is not suitable
for large bandwidths where components depend on the frequency (mainly due to
the "skin effect"). In this case, the transfer function $\left( 1\right) $
is changed in $\left( 2\right) $ or $\left( 10\right) $ which have a very
different appearance. These formulas are very close to characteristic
functions of stable probability distributions. In this paper we have proved
that the wave propagation in a coaxial cable is equivalent to a random
propagation time. The random process \textbf{A}$_{l}$ which represents it at
the distance $l$ has particular probability laws. The one-dimensional law is
stable with parameters $\left( lm,la,\frac{1}{2},-1\right) .$ $lm$ and $la$
are the parameters of position and amplitude, $\frac{1}{2}$ is the exponent
of the law. The last parameter $-1$ is particularly remarkable because this
value is matched to the causality property of linear filters. The randomly
delayed process \textbf{Z}$_{l}$ (the set of r.v. exp$\left[ i\omega
_{0}\left( t-A_{l}\left( t\right) \right) \right] )$ can be split into two
parts. The first one (the process \textbf{G}$_{l})$ is the observed process
at the end of the cable (in the case no reflexion). We have proved that the
probability distribution of $A_{l}\left( t\right) -A_{l}\left( t-\tau
\right) $ can be chosen in the class of stable distributions so that the
second part \textbf{V}$_{l}$ is outside the studied frequency band. The
model obeys the energy balance theorem because the power of the sum \textbf{G%
}$_{l}+\mathbf{V}_{l}$ is equal to the transmitted power. To conclude, this
study could be applied to a more general framework. I have proved in other
papers that the proposed model applies in many situations of propagation:
propagation of acoustics and ultrasonics waves and also propagation of
electromatic waves, in radio, radar, laser and star light (see the
bibliography).

\section{Appendix}

\bigskip 1) Let $\mathbf{X=}\left\{ X_{n},n\in \mathbb{Z}\right\} $ be a
sequence of i.i.d. (independent and identically distributed)\ r.v. (random
variables) with c.f. (characteristic function) 
\begin{equation}
\ln \text{E}\left[ e^{-i\omega X_{n}}\right] =-\sqrt{\left\vert \omega
\right\vert }\left( 1+i\text{sgn}\omega \right) .\text{ \ \ \ \ \ \ }
\end{equation}%
We know that it is one of three stable distributions with simple probability
density $f\left( x\right) $ (with the Gauss and Cauchy distributions) \cite%
{Luka} 
\begin{equation}
\text{\ }f\left( x\right) =\left\{ 
\begin{array}{c}
\frac{1}{\sqrt{2\pi }}x^{-3/2}e^{-1/2x},x>0 \\ 
0,x<0.%
\end{array}%
\right.
\end{equation}%
The main problem with this distribution is the lack of moments, which
prevents the use of the mean-square convergence. Now, we define the r.v. $Y$
by 
\begin{equation*}
Y=\sum_{k=-\infty }^{\infty }a_{k}X_{k},a_{k}>0
\end{equation*}%
where $\mathbf{a=}\left\{ a_{n},n\in \mathbb{Z}\right\} $, is a sequence of
real positive numbers. The equality%
\begin{equation}
\begin{array}{c}
\text{E}\left[ \exp \left( -\sum_{k=m}^{n}i\omega a_{k}X_{k}\right) \right] =
\\ 
\exp \left[ -\left( \sum_{k=m}^{n}\sqrt{a_{k}}\right) \sqrt{\left\vert
\omega \right\vert }\left( 1+i\text{sgn}\omega \right) \right]%
\end{array}%
\end{equation}%
comes from $\left( 11\right) ,$ using the (mutual) independence of the $%
X_{n}.$ By application of the continuity theorem of P. Levy \cite{Luka}, we
deduce that $Y$ is defined (in the sense of the convergence in distribution)
if and only if 
\begin{equation}
\sum_{k=-\infty }^{\infty }\sqrt{a_{k}}<\infty .
\end{equation}%
Moreover, this condition allows to verify the hypotheses of the
\textquotedblleft three series theorem\textquotedblright\ of A. N.
Kolmogorov \cite{Loev}, which assures the a.s. (almost sure) existence of $%
Y. $ Using $\left( 13\right) $ 
\begin{equation}
\begin{array}{c}
\text{E}\left[ e^{-i\omega Y}\right] =e^{-\lambda \sqrt{\left\vert \omega
\right\vert }\left( 1+i\text{sgn}\omega \right) }\text{ } \\ 
\text{\ with }\lambda =\sum_{k=-\infty }^{\infty }\sqrt{a_{k}}.%
\end{array}%
\end{equation}%
Consequently, the probability distributions of $Y$ and $\lambda ^{2}X_{n}$
are the same$.$

2) Now, we define the real random process \textbf{U}$_{h}=\left\{
U_{h}\left( t\right) ,t\in \mathbb{R}\right\} $ by 
\begin{equation}
U_{h}\left( t\right) =h^{2}\sum_{k=-\infty }^{\infty }\theta \left(
t-kh\right) X_{k}
\end{equation}%
for some positive function $\theta \left( t\right) $ symmetric and
decreasing on $\mathbb{R}^{+}$ . From the preceeding results, \textbf{U}$%
_{h} $ is well defined for any $h>0$ when 
\begin{equation*}
\sum_{k=-\infty }^{\infty }\sqrt{\theta \left( t-kh\right) }<\infty
\end{equation*}%
for any $t\in \mathbb{R}$. $U_{h}\left( t\right) $ follows the stable
distribution $\left( 15\right) $ with parameter 
\begin{equation*}
\lambda =h\sum_{k=-\infty }^{\infty }\sqrt{\theta \left( t-kh\right) }
\end{equation*}%
which can be taken arbitrarily close to 
\begin{equation}
\mu =\int_{-\infty }^{\infty }\sqrt{\theta \left( u\right) }du
\end{equation}%
when this quantity exists.

3) Moreover, if $t=hp_{h},\tau =hq_{h}>0$ where $p_{h}$ and $q_{h}$ are even
integers and $h$ arbitrarily small we have 
\begin{equation*}
U_{h}\left( t\right) -U_{h}\left( t-\tau \right) =h^{2}\left[ A_{h}-B_{h}%
\right]
\end{equation*}%
with 
\begin{equation*}
\left\{ 
\begin{array}{c}
A_{h}= \\ 
\sum_{k=0}^{\infty }\left[ \theta \left( -\frac{\tau }{2}+kh\right) -\theta
\left( \frac{\tau }{2}+kh\right) \right] X_{k+p-\frac{q}{2}}\geq 0 \\ 
B_{h}= \\ 
\sum_{k=0}^{\infty }\left[ \theta \left( -\frac{\tau }{2}+kh\right) -\theta
\left( \frac{\tau }{2}+kh\right) \right] X_{-k+p-\frac{q}{2}}\geq 0.%
\end{array}%
\right.
\end{equation*}%
Because the $X_{k}$ are (mutually) independent, $U_{h}\left( t\right)
-U_{h}\left( t-\tau \right) $ follows a probability distribution with c. f.
in the form%
\begin{equation*}
\begin{array}{c}
\text{E}\left[ e^{-i\omega \left( U_{h}\left( t\right) -U_{h}\left( t-\tau
\right) \right) }\right] =\exp \\ 
\left[ -h\sqrt{\left\vert \omega \right\vert }\left( \alpha _{h}+\beta
_{h}+i\left( \alpha _{h}-\beta _{h}\right) \text{sgn}\omega \right) \right]%
\end{array}%
\end{equation*}%
which does not depend on $t.$ Obviously $\alpha _{h}=\beta _{h}.$ When $%
h\rightarrow 0$ we have%
\begin{equation}
\begin{array}{c}
\lim_{h\rightarrow 0}\text{E}\left[ e^{-i\omega \left( U_{h}\left( t\right)
-U_{h}\left( t-\tau \right) \right) }\right] =\exp \\ 
\left[ -2\sqrt{\left\vert \omega \right\vert }\int_{0}^{\infty }\sqrt{\theta
\left( x-\frac{\tau }{2}\right) -\theta \left( x+\frac{\tau }{2}\right) }dx%
\right]%
\end{array}%
.
\end{equation}%
Consequently we have constructed a stationary process \textbf{U}$_{h}$ with
c.f. $\psi \left( \omega \right) $ and $\phi \left( \omega ,\tau \right) $
arbitrarily close to 
\begin{equation*}
\begin{array}{c}
\psi \left( \omega \right) = \\ 
\exp \left[ -2\sqrt{\left\vert \omega \right\vert }\left( 1+i\text{sgn}%
\omega \right) \int_{0}^{\infty }\sqrt{\theta \left( x\right) }dx\right]%
\end{array}%
\end{equation*}%
\begin{equation}
\begin{array}{c}
\phi \left( \omega ,\tau \right) = \\ 
\exp \left[ -2\sqrt{\left\vert \omega \right\vert }\int_{0}^{\infty }\sqrt{%
\theta \left( x-\frac{\tau }{2}\right) -\theta \left( x+\frac{\tau }{2}%
\right) }dx\right]%
\end{array}%
\end{equation}%
where $\theta \left( x\right) $ is a regular enough symmetric function
decreasing on $\mathbb{R}^{+}.$ We remark that (when lim$_{t\rightarrow
\infty }$ $\theta \left( t\right) =0$ quickly enough)%
\begin{equation}
\lim_{\tau \rightarrow \infty }\phi \left( \omega ,\tau \right) =\left\vert
\psi \left( \omega \right) \right\vert ^{2}
\end{equation}%
which shows some "independence" between $U_{h}\left( t\right) $ and $%
U_{h}\left( t-\tau \right) $ when $\tau $ is large.

4) Now, let assume that the process \textbf{Z}$_{l}$ defined by $\left(
6\right) $ and $\left( 7\right) $ is characterized by 
\begin{equation}
\left\{ 
\begin{array}{c}
\psi _{l,n}\left( \omega \right) =\exp \\ 
\left[ -2l\sqrt{\left\vert \omega \right\vert }\left( 1+i\text{sgn}\omega
\right) \int_{0}^{\infty }\sqrt{\theta _{n}\left( x\right) }dx\right] \\ 
\phi _{l,n}\left( \omega ,\tau \right) =\exp \\ 
\left[ -2l\sqrt{\left\vert \omega \right\vert }\int_{0}^{\infty }\sqrt{%
\theta _{n}\left( x-\frac{\tau }{2}\right) -\theta _{n}\left( x+\frac{\tau }{%
2}\right) }dx\right] \\ 
\theta _{n}\left( x\right) =n^{2}\theta _{1}\left( nx\right)%
\end{array}%
\right.
\end{equation}%
where $\theta _{1}\left( x\right) $ is positive, even, and decreasing on $%
\mathbb{R}^{+},$ with 
\begin{equation*}
\int_{0}^{\infty }\sqrt{\theta _{1}\left( x\right) }dx<\infty .
\end{equation*}%
The spectral power density $s_{V,l}^{n}\left( \omega \right) $ of the
"noise" \textbf{V}$_{l}$ is the Fourier transform of its autocorrelation
function (assuming lim$_{t\rightarrow \infty }$ $\theta _{1}\left( t\right)
=0$ quickly enough) 
\begin{equation*}
\begin{array}{c}
s_{V,l}^{n}\left( \omega +\omega _{0}\right) = \\ 
\frac{1}{2\pi }\int_{-\infty }^{\infty }e^{-i\omega \tau }\left[ \phi
_{l,n}\left( \omega _{0},\tau \right) -\left\vert \psi _{l,n}\left( \omega
_{0}\right) \right\vert ^{2}\right] d\tau .%
\end{array}%
\end{equation*}%
Using $\left( 9\right) $ and $\left( 21\right) $ we obtain the equalities%
\begin{equation*}
\left\{ 
\begin{array}{c}
s_{V,l}^{1}\left( \omega +\omega _{0}\right) =ns_{V,l}^{n}\left( n\omega
+\omega _{0}\right) \\ 
\int_{-b}^{b}s_{V,l}^{n}\left( \omega +\omega _{0}\right) d\omega = \\ 
\int_{-b/n}^{b/n}s_{V,l}^{1}\left( \omega +\omega _{0}\right) d\omega
\rightarrow _{n\rightarrow \infty }0.%
\end{array}%
\right.
\end{equation*}%
An increase of $n$ induces a widening of the spectral density (remember that
the total power of \textbf{V}$_{l}$ does not depend on $n).$ The power of 
\textbf{V}$_{l}$ in any interval $\left( \omega _{0}-b,\omega _{0}+b\right) $
can be made smaller than any quantity increasing $n.$ Consequently a device
centered on $\omega _{0}$ will only measure \textbf{G}$_{l}$ the harmonic
part of \textbf{Z}$_{l}$ if we assume that the parameter $n$ is large enough.

\end{document}